# Is the speed of light invariant or covariant ?

**Daniele Sasso** *

**Abstract**

According to the theory of ether light propagates with constant speed $c_o$ with respect to the absolute reference frame and with respect to any other reference frame the speed of light is covariant. According to the theory of special relativity the speed of light is invariant with respect to any reference frame. The new theory of reference frames gives a different answer to this question with the consideration of two speeds of light: the physical speed and the relativistic speed. After having examined a few negative aspects of the two main theories a few fundamentals of the new theory are expounded.

## 1. Introduction

Two recent papers[1][2] published in arXiv have discussed the question of the behaviour of light with respect to relativistic reference frames. D. Gezari[1] reports on the results of a Lunar Laser Ranging Test and concludes the speed of light is covariant and depends on the motion of the observer. Moreover Gezari claims this result implies the existence of a preferred reference frame for the propagation of light and he is inclined to think it coincides with an absolute reference frame.
J. Franklin[2] supports Einstein's postulate that claims the speed of light is invariant and independent of relative motion between observer and source. He challenges the validity of Gezari's experimental results and calculations. Gezari seems to be concordant with the classical theory of ether that implies the covariance of the speed of light and the existence of an absolute reference frame while Franklin supports fully Einstein's theory of relativity that implies the invariance of the speed of light with respect to all reference frames.
I think it is useful to represent here a third solution of the problem that is given by the Theory of Reference Frames[3], after having pointed out a few negative aspects of the two preceding solutions.

* engineer and independent researcher
  e_mail: dgsasso@alice.it

## 2. Critical considerations on the two main viewpoints

It is certain that whether the theory of the absolute reference frame or the theory of special relativity show several theoretical and experimental inconsistencies. Let us consider separately inconsistencies of the two theories.

### 2.1 Inconsistencies of the theory of the absolute reference frame

The first theoretical inconsistency of this viewpoint is that no one knows certainly what is the absolute reference frame. Its definition is philosophical and not scientific. Many supporters of this theory think it coincides with the ether but no one knows certainly what is the ether. Many definitions of ether have been given and many, at times also contradictory, physical properties have been assigned to ether but none of them has been fully satisfying an accepted. Moreover no experiment has pointed out the existence of ether. Also the ether is a philosophical and not scientific concept; in fact the great Greek philosopher, Aristotele, was the first to introduce the concept of ether that he considered the fifth substance of the universe (air, earth, water, fire were the other four substances).

The second theoretical inconsistency is that the absolute reference frame would be the only immobile system in an universe where all is in motion. In that case the absolute reference frame would have an origin that would be the centre of the universe and all would be in harmonious motion around this centre. The ancient world thought the earth was the centre of the universe. No experiment has pointed out the existence of the absolute reference frame and the centre of the universe.

The third theoretical inconsistency is that the absolute reference frame is the physical system where the light propagates with an unknown absolute speed. In that case in fact the speed of light measured by the earth observer ($c=\sim 3 \times 10^5$ km/s) is the speed of light relative to the earth reference frame and not to the absolute reference frame.

The fourth inconsistency is a consequence of Michelson-Morley's experiment (1887). In fact according to the negative result of this experiment H. Lorentz formulated a modified theory of ether in which the ether assumed strange properties. In particular in order to explain the absence of the modification of fringe of interference Lorentz proposed the concept of contraction of lengths towards the motion, already before introduced by G. Fitz Gerald (1891). Like this he thought to explain the isotropic nature of the speed of light that seemed be a consequence of M.M.'s experiment. Instead of clarifying the confused theoretical situation that proposal complicated further the state of theoretical physics because each time other strange properties were awarded to the ether in order to explain new phenomena.

It is clear all these inconsistencies involve a very difficult acceptance of the theory of the absolute reference frame.

## 2.2 Inconsistencies of the theory of special relativity

The first theoretical inconsistency of special relativity is that it is founded on the same model mathematical of the modified theory of the absolute reference frame: Lorentz's transformations of space-time. Lorentz reached these equations in empirical way with succeeding approximations, Einstein instead demonstrated mathematically them but on this proof there are many dubious and many authors have point out the existence of errors[4][5][6]. I note here that Einstein proved the vector composition of speeds was not valid but he used this composition in his proof. Moreover it is indeed strange that two theories, so different in their theoretical fundamentals, make use of the same mathematical model. Lorentz used these transformations in order to save the ether and the absolute reference frame. Einstein claimed that the concepts of ether and absolute reference frame were unnecessary and used those same transformations in order to save the principle of relativity.

The second theoretical inconsistency is the kinematic relativity of time based on the concept of simultaneity of moving observers. In fact it is known the simultaneity of two simulaneous events[3] can flag not only with respect to a moving observer but also to a still observer placed in not symmetric way (fig.1).

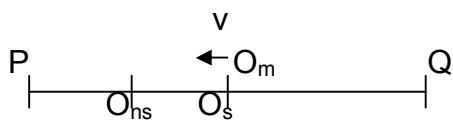

Figure 1. Two rays of light leave the points P and Q simultaneously. The still observer $O_s$ placed in symmetric way with respect to the two points claims the two rays are simultaneous. The moving observer $O_m$ claims the two rays are not simultaneous because he first sees the ray leaving the point P. The same result is obtained by the still observer $O_{ns}$ placed in not symmetric way with respect to the two points.

The two rays are simultaneous for working hypothesis and therefore the observer $O_s$ measures a true result, the two observers $O_m$ and $O_{ns}$ obtain a wrong result. We cannot claim that all the observers are equivalent similarly, the moving observer sees the two rays are not simultaneous like the not symmetric still observer but their viewpoints are wrong.

Similarly the moving observer and the not symmetric still observer can see that two events are simultaneous also when they are not simultaneous.

These reasonings show simultaneity and not simultaneity are not kinematic properties of space-time but depend on the physical state of observer.

The third theoretical inconsistency is the kinematic relativity of space. With respect to a still observer a rigid rod has a different length when it is still or when is in motion. Moreover its length changes according to its speed. Like this a rigid rod is subjected to an "accordion effect" according to its speed.

It is useful to remember these effects of space-time relativity are derived exclusively from Lorentz's transformations but they don't have any physical explanation.

The fourth theoretical inconsistency is the contradiction inside the theory of Special Relativity regarding the invariance of the speed of light.

To that end let us consider the fig.2 where A' and B' are two points inside the moving reference frame S'. A ray of light leaves A' at time 0 and reaches B' at time T' with respect to the S' observer.

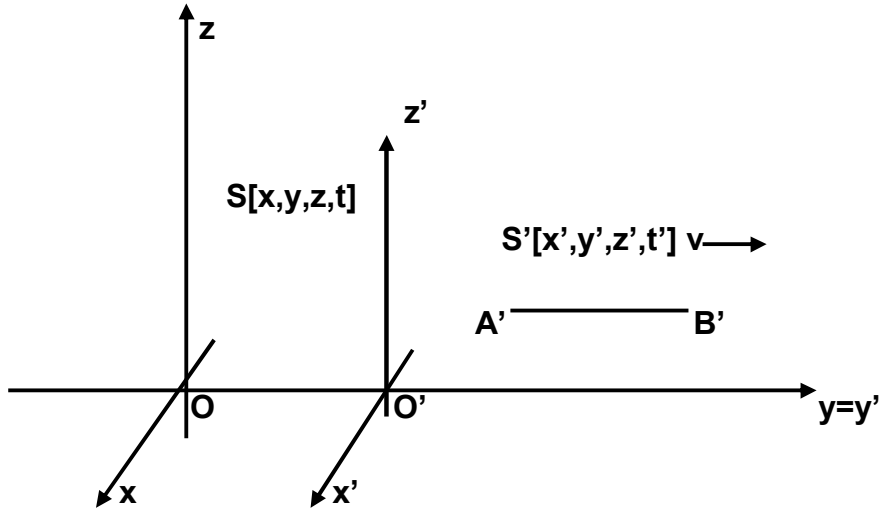

Figure 2. S[x,y,z,t] is the reference frame at rest and S'[x',y',z',t'] is the moving reference frame with speed v with respect to S.

The speed of light with respect to the reference frame S' is

$$c' = \frac{A'B'}{T'} \qquad (1)$$

With respect to reference frame at rest S in the Special Relativity the distance A'B' undergoes a contraction and has length

$$AB = \gamma\, A'B' \qquad (2)$$

with $\gamma = \sqrt{1 - v^2/c^2}$ and AB<A'B'. For the same reason always with respect to reference frame at rest S the time T' undergoes a dilation and has duration

$$T = \frac{T'}{\gamma} \qquad (3)$$

with T>T'. The speed of light with respect to the reference at rest S is therefore

$$c = \frac{AB}{T} = \gamma^2 \frac{A'B'}{T'} = \gamma^2\, c' \qquad (4)$$

with c < c'. The relationship (4) proves the postulate of invariance of the speed of light isn't respected and this contradiction is inside the theory of Special Relativity.

### 3. Theoretical fundamentals of the Theory of Reference Frames[3]

The first theoretical basis is the Principle of Reference that defines a new methodological and operating procedure in order to analyse the behaviour of physical events and specifically the light. This principle claims:

*" a physical event must be firstly analysed by a placed in symmetric way observer who is inside the reference frame tied to the physical system where the event happens"*.

The first consequence of the principle of reference is the existence of a preferred reference frame and a preferred observer. But this preferred reference frame in the theory of reference frames (briefly TR) has nothing in common with the absolute reference frame of the theory of ether. In fact in TR the preferred reference frame isn' t only because it depends on the physical system where the event happens and therefore it can change with the physical system. Moreover there is a preferred observer and he coincides with the still observer inside the preferred reference frame who is placed in symmetric way with respect to the analysed physical event. The principle of reference defines that with respect to the figure 1 the observer $O_s$ is the preferred observer who measures the correct result of the experiment and the other observers ($O_{ns}$ and $O_m$) for the analysis of their measurements must consider the result of the preferred observer and bring the corrections when it is necessary.

The second theoretical basis of the TR is the Principle of Relativity that in TR has the same meaning by Galileo and Einstein:

*" the laws, through which the states of physical systems change, are independent if these changes are attributed to either of two reference frames between them in a relative uniform motion".*

A painstaking analysis of the scientific work by Galileo and Einstein shows however that their reference frames are different. In fact Galileo in his noted work " Dialogue concerning the two chief world systems" (1632) says:

*" Shut yourself with a few friends in the gratest room that is below decks of some large ship and here observe with diligence......."* .

It is clear that Galileo considers a reference frame that is closed, isolated and not interacting with the universe. On the contrary Einstein considers open and interacting reference frames such as star reference frames. However in TR the Principle of Relativity is valid for both reference frames and it is possible to define and demonstrate a new kinematics based on the two principles before enunciated and on the concept of "inertial time" that is demonstrated[3] to be the common time for all the inertial reference frames. The inertial time is different from the absolute time of the classical physics that is associated to the absolute reference frame.

With respect to the fig.2 supposing that at time t=t'=0 O'=O, the equations of space-time transformation are in TR:

x=x'+vt' , y=y' , z=z' , t=t'   and inversely   x'=x-vt , y'=y ,  z'=z ,  t'=t        (5)

where t=t' is the inertial time. The general equations of space-time transformation are demonstrated in [3] and [7].

Let us consider now the propagation of light with respect to two reference frames in relative inertial motion. Let us consider in fig.3 Galilean reference frames and in fig.4 Einsteinian reference frames.

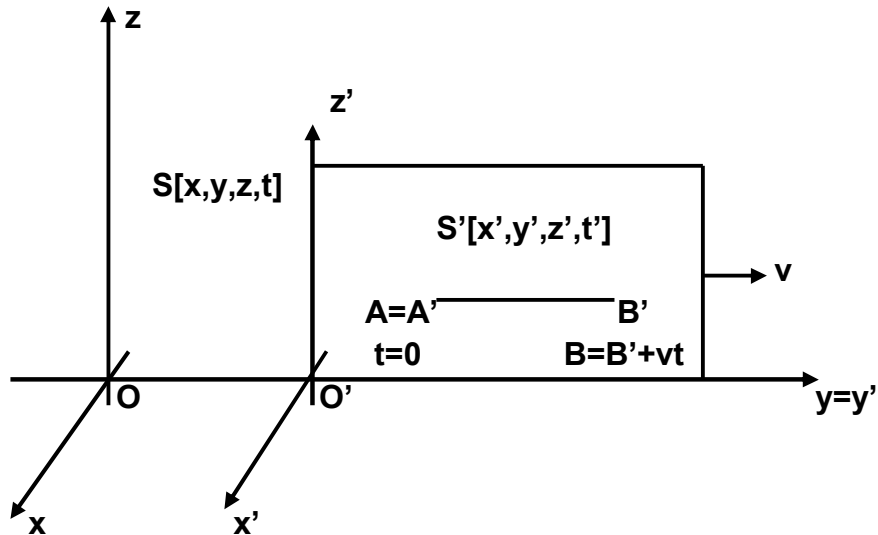

Figure 3. S[x,y,z,t] is the reference frame at rest and S'[x',y',z',t'] is the moving Galilean reference frame with uniform speed v with respect to S. S' is closed and isolated. A' and B' are still points of the moving reference frame S' and A and B are the equivalent points of the reference frame at rest S. At time t=t'=0 the origins O and O' coincide, the points A and A' coincide like the points B and B'. At the same time the ray of light leaves A', in succeeding times the point B' moves with speed v with respect to S and therefore B=B'+vt.

In the physical situation of fig.3 light propagates from A' to B' in the time T' inside the moving reference frame S' that is the preferred reference frame. The preferred observer S' measures the speed of light and finds

$$\frac{A'B'}{T'} = c_{S'} = c_o \qquad (6)$$

$c_o$ is the constant speed of light measured in all the experiments where the measure equipment is inside the reference frame.

With respect to reference frame at rest S we have

$$c_S = \frac{AB}{T} = \frac{A'B' + vT}{T}$$

Because the inertial time is the same in the two reference frames we have T'=T and therefore

$$c_S = c_o + v \qquad\qquad (7)$$

$c_o$ is the constant speed of light measured inside every reference frame and it is the "physical speed" of light. The physical speed of light is invariant with respect to any reference frame and depends on medium where light propagates through the refractive index.

$c_S$ is the "speed relativistic" of light and it depends on the speed v. The relativistic speed of light is covariant with respect to moving reference frames.

Let us consider now Einsteinian reference frames like in fig.4.

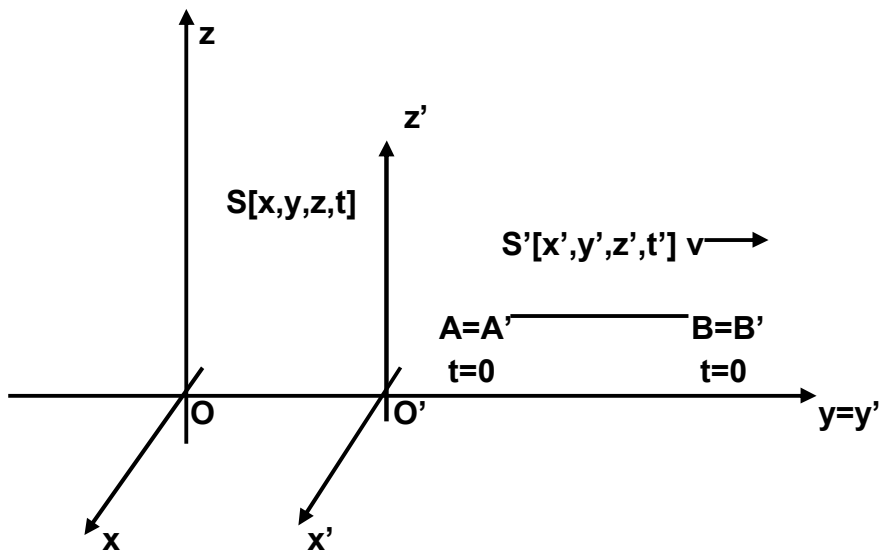

Figure 4. S[x,y,z,t] is the reference frame at rest and S'[x',y',z',t'] is the moving Einsteinian reference frame with speed v with respect to S. S' is open and interacting. At time t=t'=0 the points O', A' and B' coincide with the still points O, A and B. The light propagates in the reference frame at rest S, at time t=t'=0 it leaves the point A=A' and reaches the still point B at time T. At time T the point A' is in A+vT and therefore for the observer S' the distance covered by light at time T is A'B=AB-vT.

The source of light is in A' but because the moving reference frame is open the propagation of light happens in the reference frame at rest S. At time t'=0 the source in A' emits light but in succeeding times the propagation of light happens in S. Because the propagation of light happens in S the preferred observer is now the observer in S.

Light leaves the point A'=A at time t=t'=0 and reaches with respect to reference frame S at time T the still point B that coincides with B' at time t=t'=0. The preferred observer S measures the speed of light and finds the invariant physical speed of light

$$\frac{AB}{T} = c_S = c_o \qquad\qquad (8)$$

It is interesting to note that the result here obtained and represented by (8) is concordant with Einstein's second postulate that asserts the speed of light with respect to the reference frame at rest is independent from the speed of source. With respect to the moving reference frame S', being T'=T, the observer S' finds the covariant relativistic speed of light

$$c_{S'} = \frac{A'B}{T'} = \frac{AB - vT}{T} = c_o - v \qquad (9)$$

## 4. Considerations on the Lunar Laser Ranging Test (LLRT)

The TR claims there are two speeds of light: the physical speed and the relativistic speed. The physical speed is invariant and constant, it is the same for every reference frame where light propagates and coincides with the universal constant. The relativistic speed is covariant and depends on the relative speed between source and observer according to the vector composition of speeds.
The result of the LLRT is compatible with the Theory of Reference Frames with regards to the relativistic speed of light. The earth reference frame and the moon reference frame separately are Galilean reference frames but the earth-moon global system is an Einsteinian reference frame because it is open and interacting. In fact in the LLRT the light is emitted by a source placed on the surface of the earth and is returned by a retro-reflector on the moon. Between earth and moon there are several relative motions: the motion of rotation of the earth, the motion of rotation of the moon and the orbital motion of the moon around the earth.
Moreover the speed of motions of rotation depends on the latitude of the laboratory, at the equator the speed of the earth around its axis is 450 m/s .
It is desirable that other experiments will be performed in order to measure the relativistic speed of light also if we know the great difficulty in executing these experiments.